\documentclass[letter]{IEEEtran}

\usepackage{graphicx, dblfloatfix}
\usepackage{amsmath}
\usepackage{amssymb}
\usepackage{graphics}
\usepackage{cite}
\usepackage{amsfonts}
\usepackage{textcomp}
\usepackage{multicol}
\usepackage{multirow}
\usepackage{color}
\usepackage{fixltx2e}
\usepackage{epstopdf}
\usepackage{bm}
\usepackage{siunitx}
\usepackage{xurl}
\usepackage[dvipsnames]{xcolor}
\usepackage{multirow}
\usepackage{tabu}
\usepackage{longtable}

\begin{document}

\title{An Analog Signal Processing EIC-PIC Solution for Coherent Data Center Interconnects}

\author{Shivangi Chugh, Rakesh Ashok, Punit Jain, Sana Naaz, Aboobackkar Sidhique, and Shalabh Gupta\vspace{-10pt}

 
 \thanks{The authors are with the Department of Electrical Engineering, Indian Institute of Technology Bombay, 
 Mumbai -- 400076, India (email: shivangichugh@iitb.ac.in; rakesh.ashok@iitb.ac.in; punitjain@iitb.ac.in; sananaazsana@ee.iitb.ac.in; 
 aboobackkar@iitb.ac.in; and shalabh@ee.iitb.ac.in).
 S. Chugh and R. Ashok have contributed equally to this paper.
}}

\maketitle

\begin{abstract}

Data center interconnects (DCIs) will have to support throughputs of 400\,Gbps or more per wavelength in the near future. To achieve such high data rates, coherent modulation and detection is used, which conventionally requires high-speed data conversion and signal processing in the digital domain. Alternatively, high-speed signal conditioning and processing could be carried out in co-designed photonic and electronic integrated circuits, in the optical and electrical analog domains, respectively, to achieve reduced power consumption, latency, form factor, and cost.
A few demonstrations of analog domain processing electronic integrated circuits (EICs), including those of equalizer and carrier phase recovery (CPR) modules showcase progress in this direction in the literature. In this brief, for the first time, we present integration of a silicon photonic integrated coherent receiver (ICR) module with a CPR module, as a part of a complete coherent receiver solution. A phase shifter in the ICR (fabricated in a 220\,nm silicon-on-insulator technology) receives feedback from a CPR EIC, and the combination compensates for the time varying phase offset between the modulated signal and the unmodulated carrier in the closed loop configuration. In this proof-of-concept demonstration, we present experimental results obtained from the stand-alone silicon photonic ICR along with 
its system level integration with CPR chip, for QPSK signals. 
The technique can be extended to a higher-order modulation format, such as 16-QAM, for data rate scaling.
The proposed scheme is suitable for homodyne systems, such as polarization multiplexed carrier based self-homodyne links.

\end{abstract}

\begin{IEEEkeywords}
  
  Silicon photonic integrated circuits, integrated coherent receiver, analog signal processing, SiGe integrated circuits, phase shifter, phase detector.
  
\end{IEEEkeywords}

\IEEEpeerreviewmaketitle

\section{Introduction}
There has been a growing interest towards using coherent modulation and detection in data center interconnects (DCIs) for supporting 
the ever growing Internet traffic \cite{Nambath_JLT2020,Perin_JLT2021,Hirokawa_JLT2021}, due to the
inherent advantages of the coherent techniques \cite{Kikuchi_JLT2016}. 
Even though the coherent links possess a multitude of advantages, they also pose major challenges associated with dispersion, laser phase/frequency 
offsets, and non-zero laser linewidth. The impairments due to these effects are conventionally overcome using digital signal processing. However,  the severity of the impairments introduced in the DCIs is not as significant as that in the long-haul links. Therefore, the power hungry data conversion and digital signal processing based approach can be avoided and substituted by optical domain processing and analog signal processing (ASP) for considerable
power, complexity, form-factor, and latency reduction \cite{Ashok_JLT2021}. 
The enhanced energy efficiency and reduced complexity in the ASP approach are attributed to
the omission of analog to digital converters (ADCs) and massive number of operations in digital signal processors (DSPs).  
In the literature, all-analog coherent receiver solutions compensating for dispersion and laser offsets are presented in 
\cite{Nambath_JLT2020,Hirokawa_JLT2021,Verplaetse_JSSC2020,Ashok_JLT2021,Ashok_JLT2021_1}. When projected to the 400ZR standard, 
the ASP approach implemented in a 7\,nm FinFET technology can save $\sim$3.6\,W compared to its digital counterpart \cite{Ashok_JLT2021}.


Major requirements for next-generation DCIs include high power efficiency, high bandwidth, high reliability, small form factor, low latency, and low cost. A photonic integrated circuit (PIC) comprising optical components along with electronics integrated on a chip can make the entire solution very efficient. Monolithic integration on silicon is lately gaining popularity,
mainly due to low-cost manufacturing and maturity of CMOS fabrication processes \cite{Jalali_JLT2006}. The combination of a silicon photonic integrated circuit (SiPIC) with analog electronic integrated circuits (EICs) is expected to bring down the form-factor, power consumption, and cost of transceivers in coherent DCIs. 

Earlier, a 16-QAM coherent DCI transmitter demonstrated in \cite{Ahmed_JSSC2020}, has an integrated PIC designed in silicon-on-insulator 
(SOI) with EIC designed in SiGe platforms 
supports 552\,Gb/s/$\lambda$. A 56\,Gb/s SiPIC transmitter 
in 3D-Integrated PIC25G and ST-55\,nm BiCMOS
technology is presented in \cite{Temporiti_ISSC2016}.
An SiPIC coherent receiver comprising all optical components and trans-impedance 
amplifiers (TIAs) on the same chip for dual-polarization QPSK links has been demonstrated in \cite{Doerr_JLT2010,Tsunashima_OE2012}. 
A quasi-coherent optical receiver, designed in 130\,nm SiGe BiCMOS technology with major components being PIN detectors, TIAs,
envelope detectors, and limiting amplifiers for 28\,Gbaud has been demonstrated in \cite{Valdecasa_TCAS22021}.

\begin{figure}[tb!]
	\centering{
		
		\hspace{0cm}\includegraphics[width=.49\textwidth]{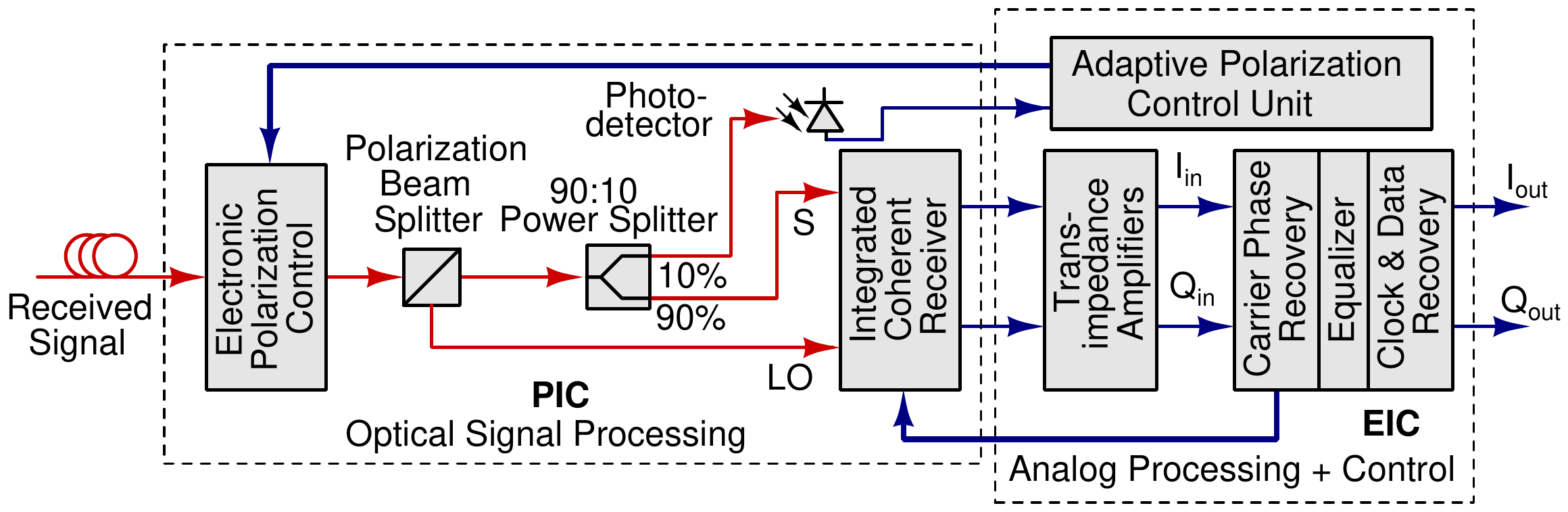}
}
		\caption{ Conceptual electronic and photonic integrated analog signal processing solution block diagram for 
		polarization multiplexed carrier based self-homodyne link. PIC: Photonic integrated circuit; EIC: Electronic 
		integrated circuit;
		S: Modulated signal; and LO: Unmodulated carrier.}		
		\label{fig:monolithic}
	\end{figure}

This work presents the demonstration of a SiPIC integrated coherent receiver (ICR) in a Costas loop with carrier phase recovery (CPR) chip (the stand-alone CPR chip has been presented earlier in \cite{Ashok_JLT2021}). 
Functional validation of the EIC-PIC receiver has been performed for the QPSK homodyne link.
This Costas loop aims at removing the time varying phase offset between the modulated 
and local oscillator signals, for reliable recovery of data in coherent homodyne systems, especially in a polarization multiplexed carrier based
self-homodyne (PMC-SH) links \cite{Kamran_JLT2020}. The EIC-PIC ASP based coherent receiver exhibits an essential step in the direction of achieving highly efficient DCIs.

\section{System design and overview}

The proposed conceptual ASP receiver architecture that can be monolithically integrated for providing a complete solution for PMC-SH based DCI 
is shown in Fig. \ref{fig:monolithic}. A generic ASP receiver architecture for polarization division multiplexed system with laser phase/frequency offset 
is detailed in \cite{Ashok_JLT2021,Nambath_JLT2020}.
The modulated signal ($S$) and unmodulated carrier ($LO$) from the received signal are separated with the aid of an adaptive polarization controller consisting of an electronic polarization controller, 
polarization beam splitter, power splitter, photodetector, 
and adaptative polarization control circuit \cite{Kamran_JLT2020}. 
The polarization segregated optical signals $S$ and $LO$ are given to the ICR. 
The electrical signals from the ICR output after the TIAs
are given to the analog processing and control unit. 
Laser offsets, dispersion, and clock misalignments are corrected using CPR, equalizer, and clock and data recovery circuits, respectively, 
which are the major constituents of the analog processing and control unit. 
For longer channel lengths, equalization can be used prior to CPR for correcting the massive dispersion introduced.

	\begin{figure}[t!]
    \centering
    
    \includegraphics[width=.47\textwidth]{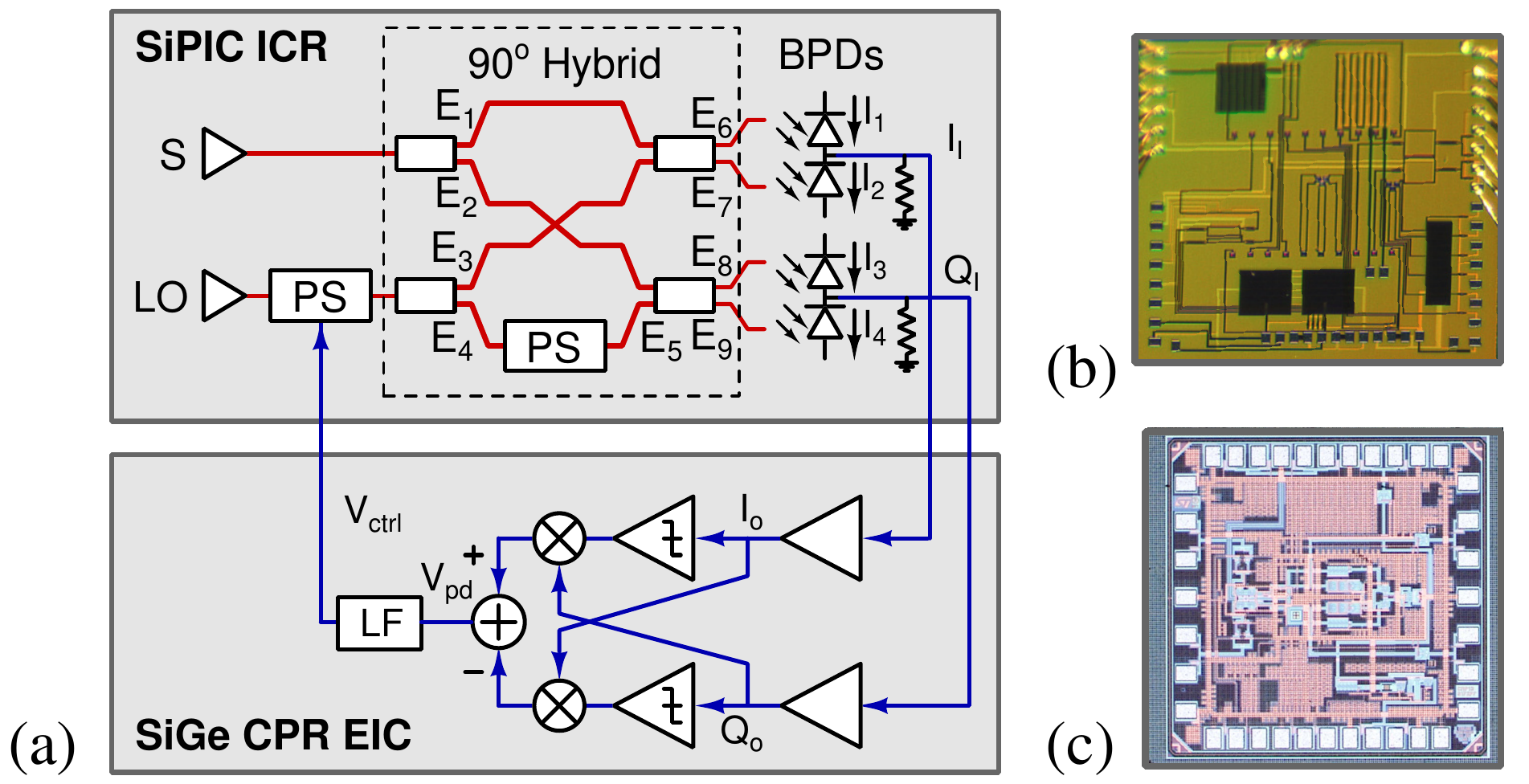} \\
        \caption{(a) Simplified block diagram of the key components of the proposed coherent receiver solution presented in this work;
        Micrographs of application specific integrated circuits: (b) Silicon photonic integrated circuit; and
    (c) SiGe BiCMOS analog domain carrier phase recovery chip
        PS: Phase shifter; BPD: Balanced photodetector; LF: Loop filter; 
         $S$: Modulated signal; $LO$: Unmodulated carrier; $I_I$, $Q_I$: 
        Input in-phase and quadrature-phase signals; $I_O$, $Q_O$: 
        Output in-phase and quadrature-phase signals; $V_{pd}$: Phase detector output; and $V_{ctrl}$: Control voltage.}
    \label{fig:EPIC_architecture}   
  \end{figure}
  
The architecture of the ASP coherent receiver is shown in Fig. \ref{fig:EPIC_architecture}. 
The system comprises an SiPIC ICR and an SiGe analog domain CPR EIC. 
The ICR receives a modulated signal ($S=\exp\{j[\omega_st+\phi_m(t)+\phi_{s}(t)]\}$) and an unmodulated carrier ($LO=\exp\{j[\omega_{lo}t+\phi_{lo}(t)]\}$) at the S and LO ports, 
respectively. 
Here, $\omega_s$, $\phi_m$, $\phi_s$, $\omega_{lo}$, and $\phi_{lo}$ denote frequency of the carrier laser, message phase, 
random phase fluctuations of the carrier laser, frequency of the LO laser, and random phase fluctuations of LO laser, respectively.
These signals are given to the 90$^{\circ}$ optical hybrid and electric fields at various nodes of 90$^{\circ}$ optical hybrid are $E_1=S/\sqrt{2}$, $E_2=-jS/\sqrt{2}$, $E_3=-jLO/\sqrt{2}$, 
 $E_4=LO/\sqrt{2}$, $E_5=-jLO/\sqrt{2}$, $E_6=(S+LO)/2$, $E_7=(S-LO)/2$, $E_8=(S+jLO)/2$, and $E_9=(S-jLO)/2$.
The outputs of the hybrid are given to the photodetectors for generating the (normalized) photocurrents 
$I_1=1+\cos[(\omega_s-\omega_{lo})t+\phi_m(t)+\phi_s(t)+\phi_{lo}(t)]$, 
$I_2=1-\cos[(\omega_s-\omega_{lo})t+\phi_m(t)+\phi_s(t)+\phi_{lo}(t)]$,
$I_3=1+\sin[(\omega_s-\omega_{lo})t+\phi_m(t)+\phi_s(t)+\phi_{lo}(t)]$, and
$I_4=1-\sin[(\omega_s-\omega_{lo})t+\phi_m(t)+\phi_s(t)+\phi_{lo}(t)]$. The balanced photodetector (BPD)
stage removes DC (from $I_1$ to $I_4$) and provides differential currents, which 
are subsequently converted into voltage signals $I_I=\cos[(\omega_s-\omega_{lo})t+\phi_m(t)+\phi_s(t)+\phi_{lo}(t)]$ and 
$Q_I=\sin[(\omega_s-\omega_{lo})t+\phi_m(t)+\phi_s(t)+\phi_{lo}(t)]$. These signals are given to the CPR chip comprising a
cross-correlator phase detector (PD) and the inputs to the PD are $I_O=\cos[\phi_m(t)+\phi_{err}(t)]$ and 
$Q_O=\sin[\phi_m(t)+\phi_{err}(t)]$, where $\phi_{err}(t)=(\omega_s-\omega_{lo})t+\phi_s(t)+\phi_{lo}(t)$ denotes the cumulative phase error 
in $I_O$ and $Q_O$. 

In the case of frequency synchronized links, such as PMC-SH links, $\omega_s=\omega_{lo}$. The 
PD output, $V_{pd}\approx\phi_{err}(t)$ for $\phi_{err}(t)\in(-\pi/4$,$\pi/4)$. The $V_{pd}$ signal is given 
to the phase shifter (PS) of SiPIC in $LO$'s path through a loop filter (LF) so that the phase delay ($\phi_d$) added to $LO$ by the PS 
is equal and opposite to $\phi_{err}(t)$. In the closed loop, $\phi_d(t)=-\phi_{err}(t)=-[\phi_s(t)+\phi_{lo}(t)]$, thus compensating for the phase error. 
The LF averages $V_{pd}$ and stabilizes the loop. Under the 
closed loop condition, at steady state, $I_O=\cos[\phi_m(t)]$ and $Q_O=\sin[\phi_m(t)]$, which are the desired message phases after phase error 
correction and are used for data recovery in further steps using equalizer and clock and data recovery circuits.

  \begin{figure}[tb!]
	\centering{
		\hspace{-.1cm}\includegraphics[width=.42\textwidth]{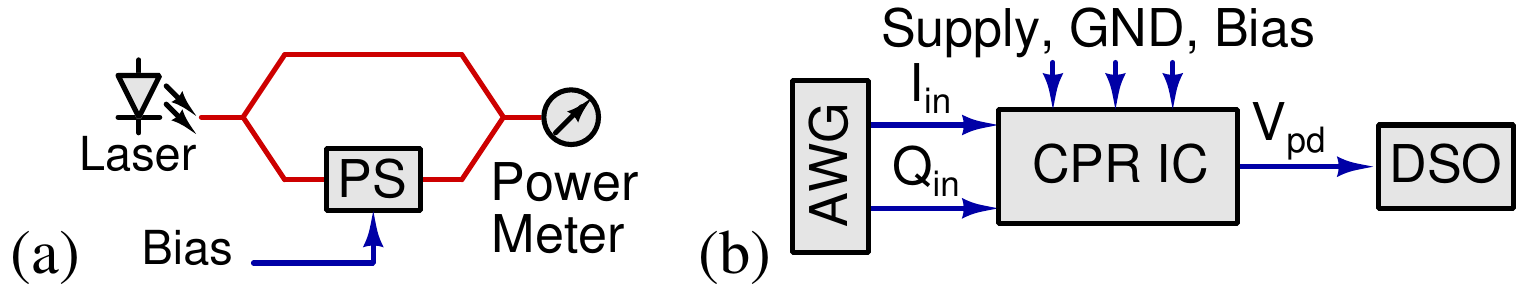}\\
}
		\caption{ Experimental characterization setups for: (a) Phase shifter in silicon photonic integrated coherent receiver; and 
		(b) Phase detector in carrier phase recovery chip.}		
		\label{fig:PMPDChar_setup}
	\end{figure}

  \begin{figure}[tb!]
	\centering{
		\begin{tabular}{cccc}
		\hspace{-.1cm}\includegraphics[width=.22\textwidth]{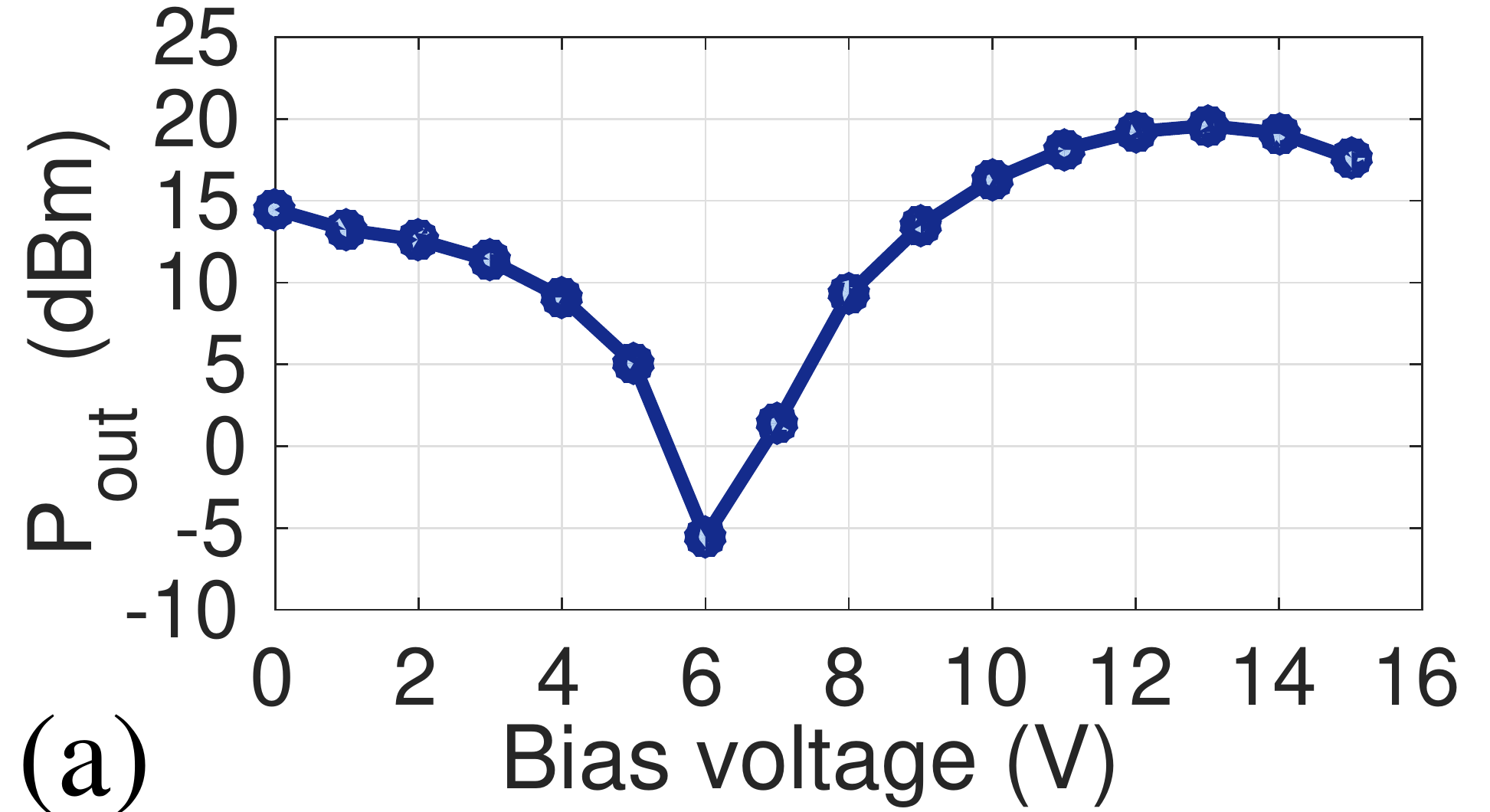}& 
		\hspace{-.15cm}\includegraphics[width=.22\textwidth]{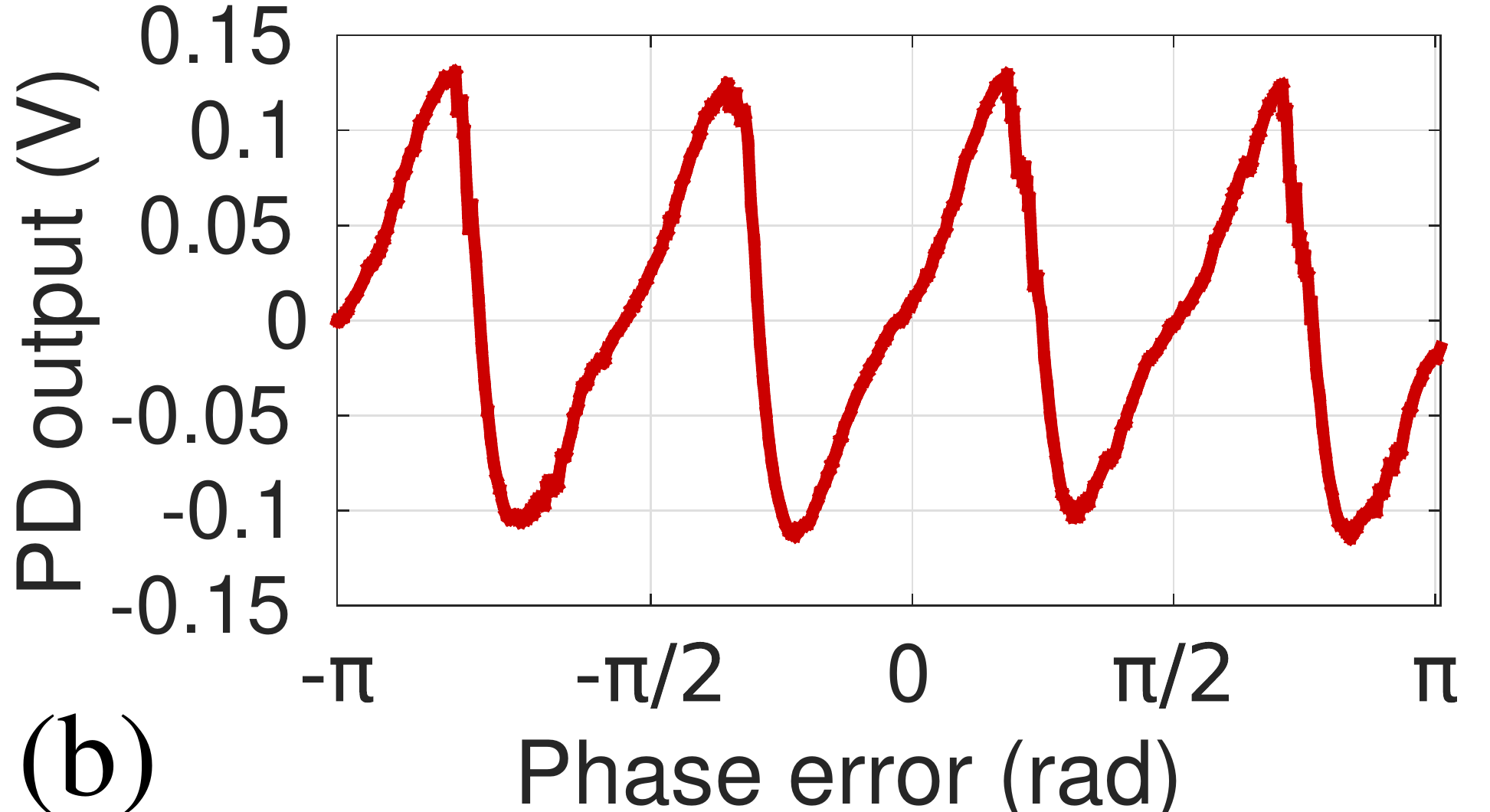}\\
\end{tabular}}
		\caption{ Experimental characterization results: (a) Phase shifter characterization using an interferometer 
		structure; and (b) Phase detector characteristics obtained using 1\,Gbaud QPSK signals with 1\,MHz frequency offset.}		
		\label{fig:SiPhCPRC_char}
	\end{figure}

  \section{EIC and PIC design}
  The presented scheme comprises two application-specific integrated circuits: an SiPIC ICR and an SiGe CPR EIC, the details of which are provided in the following 
  subsections.
  
%
  
  \subsection{SiPIC ICR chip}
  
       \begin{table}[tb!]
\centering
   \caption{Salient features of components used in SiPIC ICR chip}
\begin{tabular}{lccccc}
\hline\hline
{\hspace{-.15cm}Component} & \hspace{-.25cm}Salient features\\  
              \hline\hline
\hspace{-.15cm}VGC&\hspace{-.25cm}Insertion loss = 4\,dB, 1\,dB BW = 28\,nm\\
\hspace{-.15cm}1$\times$2 MMI& \hspace{-.25cm}Excess loss = 0.04\,dB, Power imbalance = 0.02\,dB\\
\hspace{-.15cm}Thermo-optic PS&\hspace{-.25cm}Phase change speed = 50\,kHz, $V_{\pi}$ $\simeq$ 6\,V\\
\hspace{-.15cm}Ge photodetector&\hspace{-.25cm}BW = 50\,GHz, $I_{dark}$ = 15\,nA, Responsivity = 0.45\,A/W\\
 \hline\hline
\end{tabular}
\label{tab:ICR_chip_summary}
\end{table}
  
  The SiPIC ICR is designed and fabricated in IMEC's ISIPP50G process, which is an SOI technology and is a suitable candidate for monolithic integration with EIC technologies \cite{lockwood2010silicon}. 
  The micrograph of the ICR is shown in Fig. \ref{fig:EPIC_architecture}(b). 
  The chip comprises multiple designs, among which ICR is focused in this work. Here, the SiPIC design is specific for C-band transverse electric operation. 
  Major building blocks of SiPIC ICR are vertical grating couplers (VGCs), 90$^{\circ}$ hybrid, and BPDs.
  SiPIC uses VGCs for surface coupling of optical signals to the chip. 
  The PS, an important section of the ICR, is characterized using an interferometer structure shown in Fig. \ref{fig:PMPDChar_setup}(a). 
  The PS has a
  half-wave voltage ($V_{\pi}$) of $\sim$6\,V, corresponding to the minimum output power,
  as shown in the characterization result in Fig. \ref{fig:SiPhCPRC_char}(a). 
    The 90$^{\circ}$ optical hybrid is made up of two 1$\times$2 multi-mode interferometers (MMIs), a PS, and two 2$\times$2 MMIs.
  The salient features of the components used in the design are summarized in Table \ref{tab:ICR_chip_summary}.
          A 50\,$\Omega$ load resistor is provided at the output of each BPD pair to convert the photocurrent into voltage and to enable matching for high frequency signaling. 
  The ICR structure has two RF ports for each in-phase and quadrature-phase signal, arranged in ground-signal-ground format. 
  Supply, ground, and bias DC pads are wire bonded on a printed circuit board (PCB).

    	    \begin{figure}[t!]
    \centering
    
    \includegraphics[width=.38\textwidth]{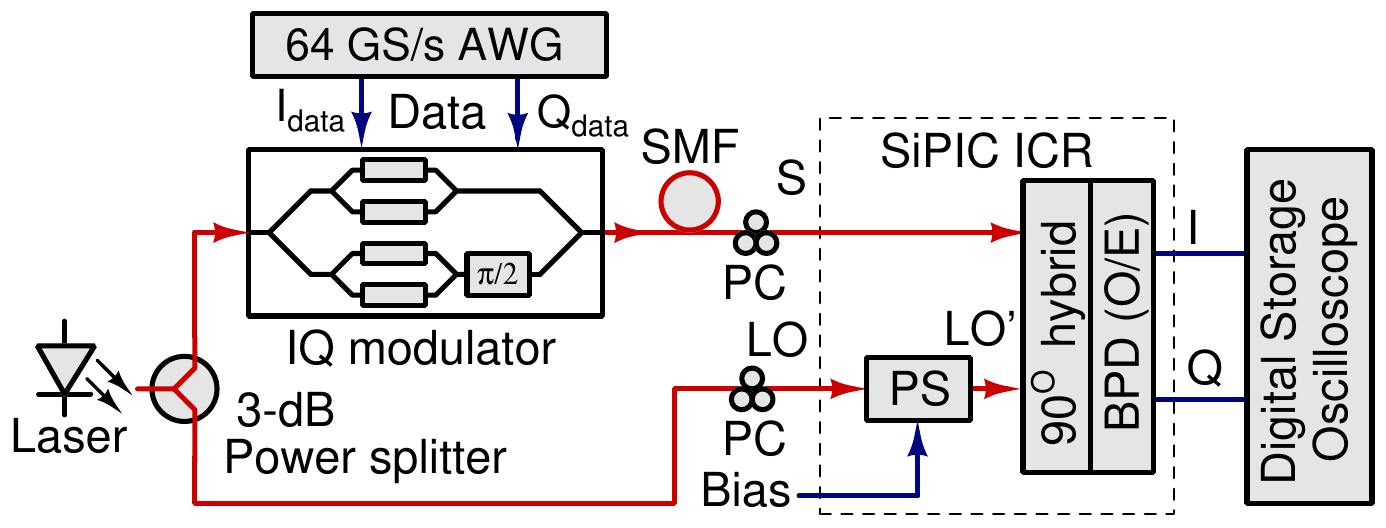} \\
    \caption{Experimental set-up for characterizing silicon photonic integrated coherent receiver. 
    AWG: Arbitrary waveform generator;
    $I_{data}$, $Q_{data}$: In-phase and quadrature-phase message signals; SMF: Single mode fiber; 
    PS: Phase shifter;
    PC: Polarization controller;
    BPD: Balanced photodetector; ICR: Integrated coherent receiver; 
    $S$: Modulated signal; and $LO$ and $LO'$: Unmodulated carrier before and after phase shifting.    
    }
    \label{fig:PICChar}   
  \end{figure}

  \begin{figure}[t!]
    \centering
    \begin{tabular}{ccc}
   \hspace{-.5cm}   \includegraphics[width=.17\textwidth]{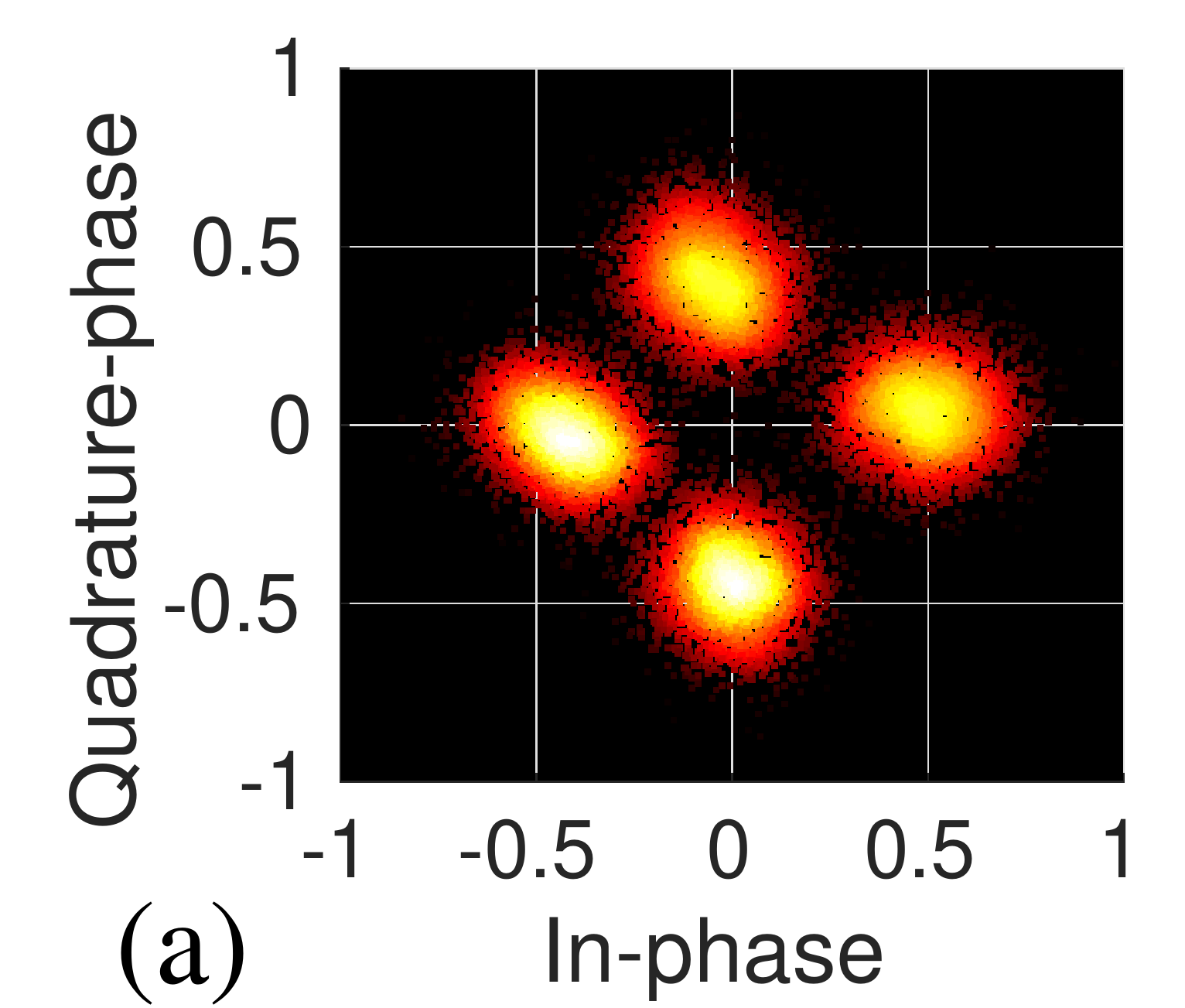} &     \hspace{-.68cm}
    \includegraphics[width=.17\textwidth]{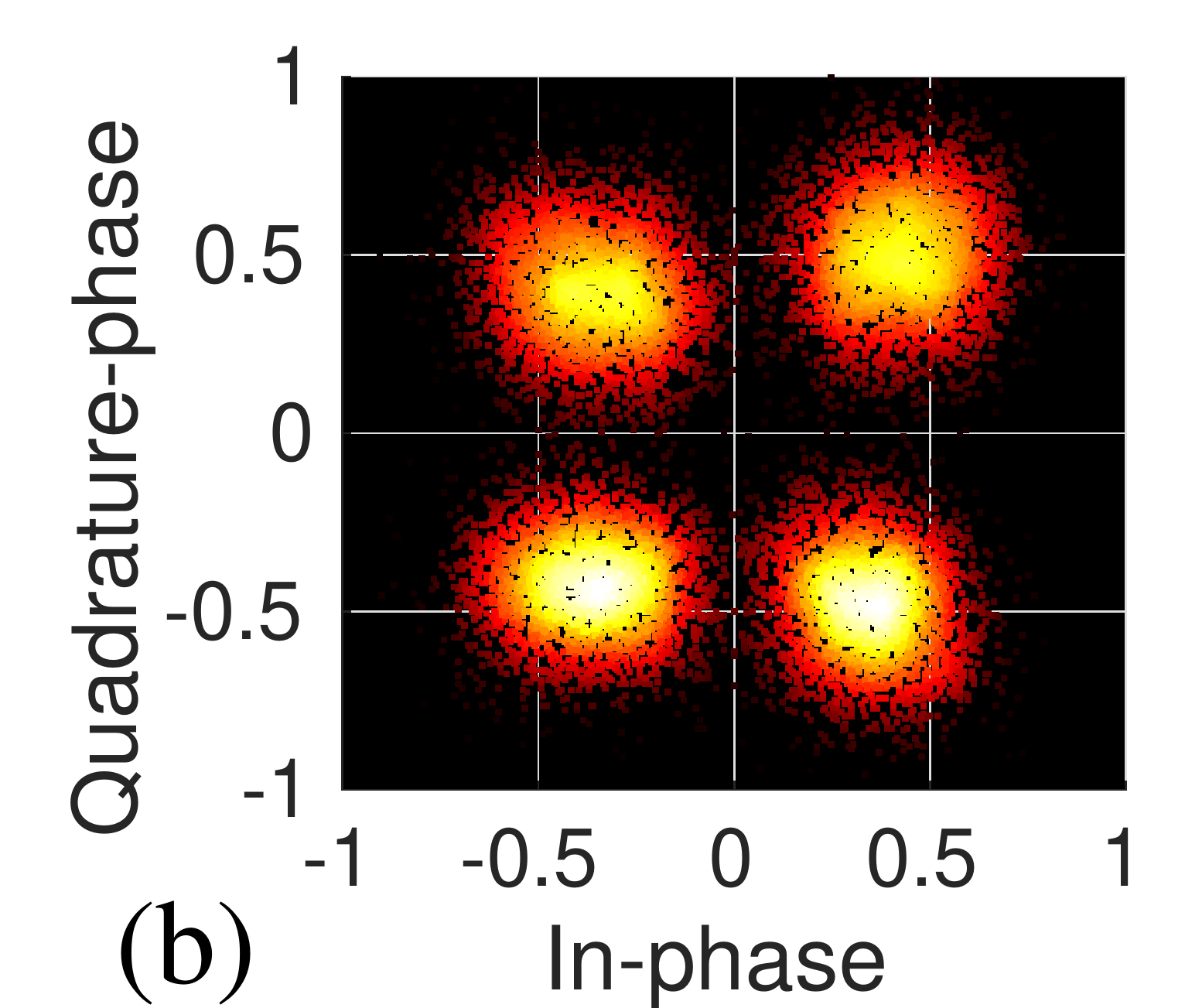}  &     \hspace{-.68cm}
     \includegraphics[width=.17\textwidth]{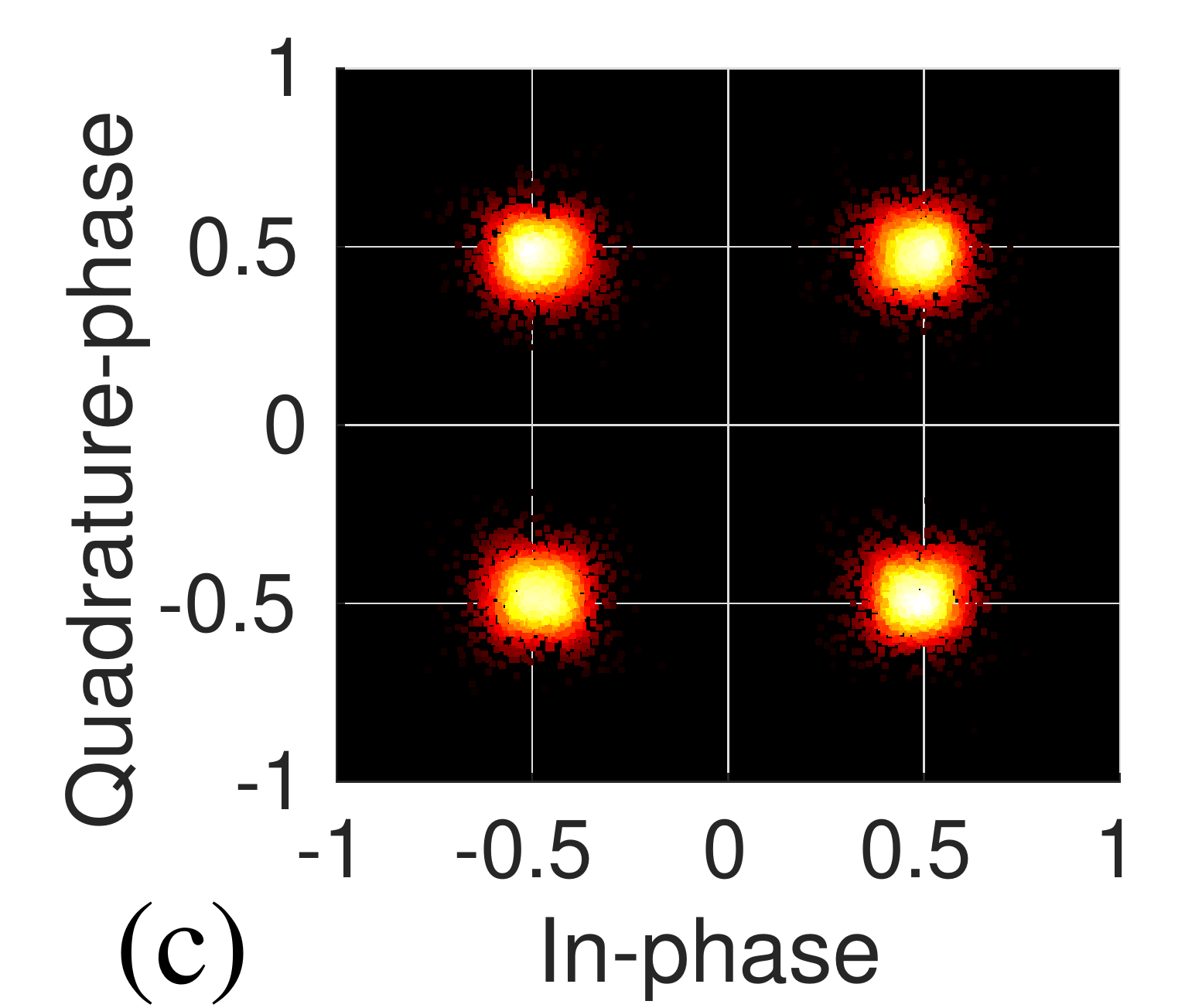}      
     \\
    \end{tabular}
    \caption{Results obtained with silicon photonic integrated coherent receiver and analog signal processing for 10\,m 8\,Gbps QPSK link.
    Constellation diagrams of I/Q signals at:
    (a) Silicon photonic integrated coherent receiver output (experimental); 
    (b) Carrier phase recovery and compensation (post-processing); and (c) Equalization (post-processing).}
    \label{fig:SiPICICR_SA}
  \end{figure}
  
  \subsubsection*{Characterization of SiPIC ICR}
  The experimental setup used for characterizing the proof-of-concept SiPIC ICR is shown in Fig. \ref{fig:PICChar}. A 1550\,nm laser output 
  is split using a power splitter, one of the outputs of which is QPSK modulated by 4\,Gbaud electrical signals generated by an arbitrary waveform generator 
  (AWG), while other acts as $LO$. The $S$ and $LO'$ signals are mixed in SiPIC ICR to generate $I$ and $Q$ electrical signals shown in 
  Fig. \ref{fig:SiPICICR_SA}(a). The PS is not biased as the feedback is not essential in this characterization.
  The data rate, in this case, is mainly constrained by the probes used to extract electrical signals from the
  ICR. The $I$ and $Q$ signals recorded through oscilloscope are subjected to post-processing through behavioral carrier phase recovery and 
  compensation and equalization in sequence and the corresponding results are shown in Figs. \ref{fig:SiPICICR_SA}(b) and (c), respectively.

       \begin{table}[tb!]
\centering
   \caption{Characterisitics of circuit blocks in SiGe CPR chip}
\begin{tabular}{lccccc}
\hline\hline
{\hspace{-.15cm}Circuit} & Key characterisitics\\  
              \hline\hline
\hspace{-.15cm}Input stage     & \hspace{-.3cm}BW = 38\,GHz, ${S_{dd11}}<-$10\,dB, $Z_{in,d}$ $\simeq$ 100\,$\Omega$ \\ 
\hspace{-.15cm}Delay cell& \hspace{-.3cm}BW = 19.2\,GHz, Gain = 1.8\,dB, Group delay = 28.1\,ps\\
\hspace{-.15cm}Limiting amplifier&\hspace{-.3cm}BW = 27.5\,GHz, Gain = 41\,dB, Group delay = 28\,ps\\
\hspace{-.15cm}Multiplier    &\hspace{-.3cm}BW = 20.89\,GHz, Gain = 2.75\,dB\\
\hspace{-.15cm}Adder    &\hspace{-.3cm}BW = 24\,GHz, Gain = 5.2\,dB\\
\hspace{-.15cm}Output  stage   &\hspace{-.3cm}BW = 36\,GHz, ${S_{dd22}} < -$10\,dB, $Z_{out,d}$ $\simeq$ 100\,$\Omega$\\ 
 \hline\hline
\end{tabular}
\label{tab:CPRC_chip_summary}
\end{table}

  \subsection{SiGe CPR chip}
  The micrograph of analog domain CPR chip designed in ST-130\,nm BiCMOS technology ($f_t$ = 230\,GHz and $f_{max}$ = 280\,GHz) is shown in Fig. \ref{fig:EPIC_architecture}(c). 
    The chip occupies an area of 1.2\,mm$\times$1.2\,mm consisting of a total number of 36 pads. 
   The chip is designed for 25\,Gbaud QPSK and 16-QAM modulation formats with all the individual circuits to 
  have a bandwidth (BW) of at least 0.7 times the desired baud rate and 400\,mV$_\text{pp}$ differential swing.
   The major constituents of this chip are SSB mixer and PD, which in turn comprise limiting amplifiers, delay cells, 
  multipliers, adders, buffers, and biasing circuits, the key characteristics of which with 50\,fF capacitive load are presented in Table \ref{tab:CPRC_chip_summary}. 
  The schematics of these differential circuits (made up of BJTs and MOSFETs), design, and simulation results are detailed in \cite{Ashok_JLT2021}.
   The cross-correlator PD generates a voltage proportional to the phase error in the input signals. 
  The PD of the CPR chip, which is the effective section of the chip used for the demonstration, consumes $\sim$412\,mW power.
  The PD is characterized using the setup shown in Fig. \ref{fig:PMPDChar_setup}(b) by feeding 1\,Gbaud QPSK 
  modulated data with a frequency offset of 1\,MHz. The PD output has saw-tooth characteristics with a phase periodicity of $\pi/2$ and gain of 0.16\,V/rad, as shown in Fig. \ref{fig:SiPhCPRC_char}(b). 
  The input and output differential high frequency signals are arranged in ground-signal-signal-ground-signal-signal-ground format with 
  $\sim$100\,$\Omega$ differential source and load impedances supporting 25\,Gbaud signaling.
  This chip's 
  functionality with discrete components and commercial InGaAs ICR has been demonstrated in \cite{Ashok_JLT2021,Ashok_OFC2019,Ashok_CLEO2020}.

  \section{Proof-of-concept demonstration}

\begin{figure}[t!]
    \centering
    
    \includegraphics[width=.495\textwidth]{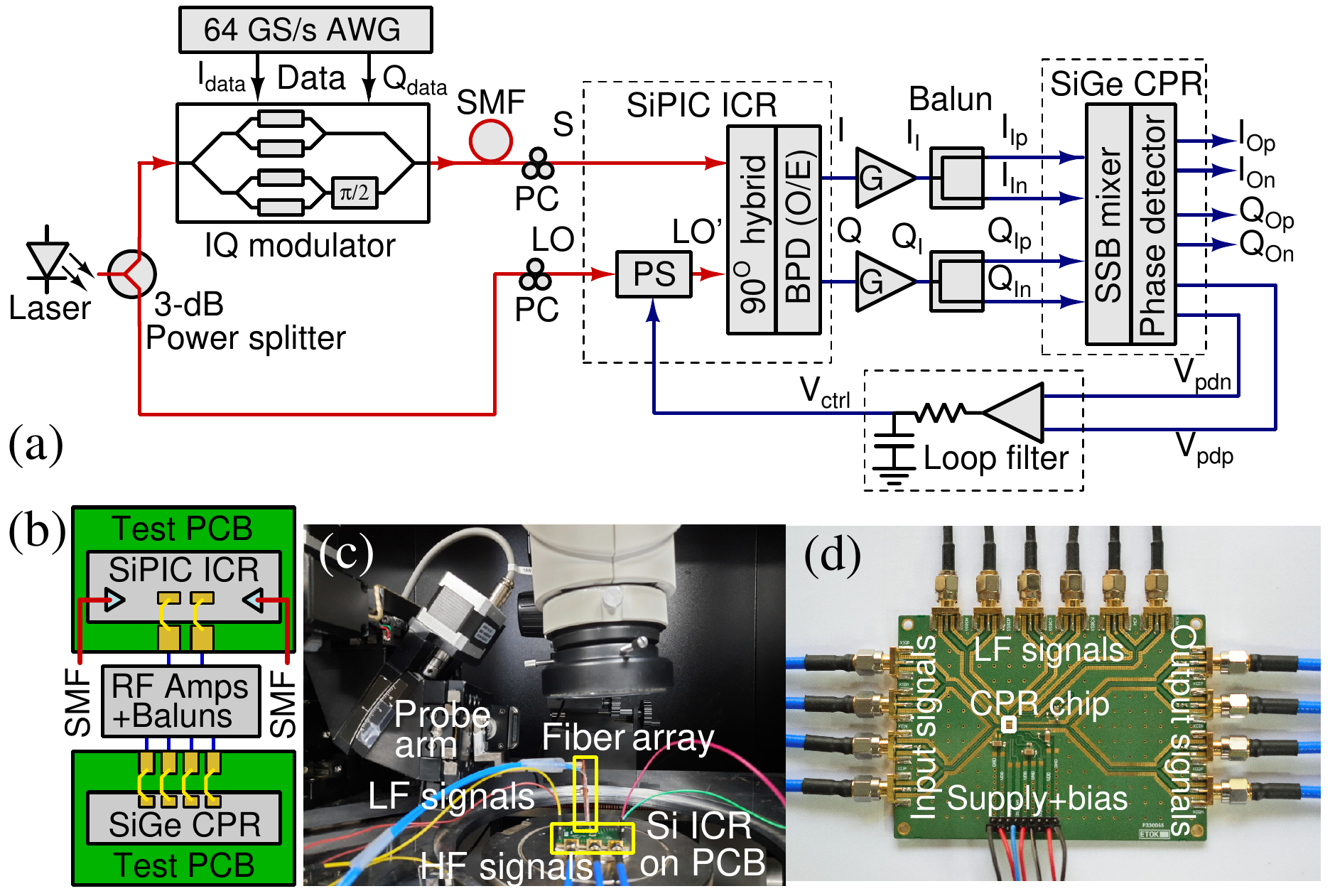} \\
    \caption{(a) Experimental set-up of ASP coherent receiver with silicon photonic integrated circuit and SiGe electronic
    integrated circuit; (b) Interfacing silicon photonic integrated coherent receiver and BiCMOS carrier phase recovery chips;    
    (c) Experimental setup of silicon photonic integrated coherent receiver chip; and
    (d) Experimental setup for SiGe analog domain carrier phase recovery chip.
    AWG: Arbitrary waveform generator;
    $I_{data}$, $Q_{data}$: In-phase and quadrature-phase message signals; SMF: Single mode fiber; 
    PS: Phase shifter;
    PC: Polarization controller; 
    BPD: Balanced photodetector; ICR: Integrated coherent receiver; CPR: Carrier phase recovery; PCB: Printed circuit board;
    $S$: Modulated signal; $LO$, $LO'$: Unmodulated carrier before and after phase shifting;    
    $I_I$, $Q_I$: 
        Input in-phase and quadrature-phase signals; $I_O$, $Q_O$: 
        Output in-phase and quadrature-phase signals; $V_{pd}$: Phase detector output; and $V_{ctrl}$: Control voltage.
    }
    \label{fig:Exp_EPIC}   
  \end{figure}

The proposed EIC-PIC ASP coherent receiver is validated for proof-of-concept, 
at 2\,GBaud with QPSK homodyne optical link. Figure \ref{fig:Exp_EPIC}(a) shows the experimental setup used for demonstrating the proposed approach. 
A C-band, continuous wave, single frequency, and external cavity laser output 
is split using a 3-dB power splitter to provide a carrier to the IQ modulator at the transmitter and an $LO$ for coherent detection 
at the receiver for ensuring no frequency offset and drift between $S$ and $LO$. 
The $I_{data}$ and $Q_{data}$ (PRBS-7) message signals generated by an AWG modulate the carrier using an IQ modulator. 
The modulated optical signal is transmitted over a 10\,m SMF channel, which results in time varying phase offset between $S$ and $LO$. 
The modulated signal's polarization is adjusted using a polarization controller (PC) and applied to the S port of the 
SiPIC ICR through VGC. The unmodulated carrier is applied to the LO port using another VGC through PC to control the polarization. 
The receiver comprises SiPIC ICR, amplifiers, balanced to unbalanced lines (baluns), SiGe CPR chip, and LF arranged in a feedback loop structure.
$LO$ is given to the hybrid through a PS to control its phase in the closed loop. 
The PS in the hybrid is biased to provide a phase shift of $\pi/2$ to the $LO$. 
Outputs from the hybrid are given to the BPD pairs, biased with --1\,V and 1\,V supply to generate photocurrents comprising details of message phase, message amplitude, and cumulative phase error. 
Interconnection of SiPIC ICR and SiGe CPR with their corresponding test PCBs is illustrated in Fig. \ref{fig:Exp_EPIC}(b).

The SiPIC experimental setup on a probe station is shown in Fig. \ref{fig:Exp_EPIC}(c).
The optical signals are coupled to the ICR through a fiber array while the electrical signals are wire-bonded on SiPIC PCB. 
 The coupling of the $S$ and $LO$ optical signals from the fiber array is controlled through a mechanical probe arm, which is monitored manually. 
The in-phase ($I$) and quadrature-phase ($Q$) outputs of SiPIC have a swing of $\sim$5\,mV$_\text{pp}$ with corresponding 
maximum BPD photocurrents of $\sim$300\,\si\micro A.

The electrical $I$ and $Q$ signals obtained at the output of SiPIC ICR are amplified (using an amplifier of gain $\simeq$ 25\,dB) and then converted into differential signals using baluns
(insertion loss $\simeq$ 3\,dB) that give out unbalanced outputs, which are 3\,dB below both
$I$ and $Q$ signals. The amplification and balanced to unbalanced conversion are essential as the CPR chip requires a minimum differential swing of 50\,mV$_{\text{pp}}$ at its inputs for its operation.

Figure \ref{fig:Exp_EPIC}(d) shows the experimental setup of the CPR chip. The aluminum pads of the CPR chip are (gold) wire bonded to
electroless nickel immersion gold tracks on a four-layer PCB
fabricated on an FR-4 substrate. The high frequency input/output signals are fed/extracted through sub-miniature version-A connectors while supply, ground, and bias connections are provided through berg-stick connectors.
The differential signals $I_I$ ($=I_{Ip}-I_{In}$) and $Q_I$ ($=Q_{Ip}-Q_{In}$) 
are fed to the CPR chip. The chip gives out 
$I_O$ ($=I_{Op}-I_{On}$), $Q_O$ ($=Q_{Op}-Q_{On}$), and $V_{pd}$ ($=V_{pdp}-V_{pdn}$). The $V_{pd}$ signal is given to 
the PS through the LF, which translates the voltage levels of $V_{pd}$ to voltage levels required for PS, 
averages $V_{pd}$, and stabilizes the loop. 
The voltage waveforms at the CPR chip output are recorded in open loop and closed loop conditions.

	  \begin{figure}[tb!]
	\centering{
		\begin{tabular}{ccccccc}
		\hspace{-.05cm}\includegraphics[width=.16\textwidth]{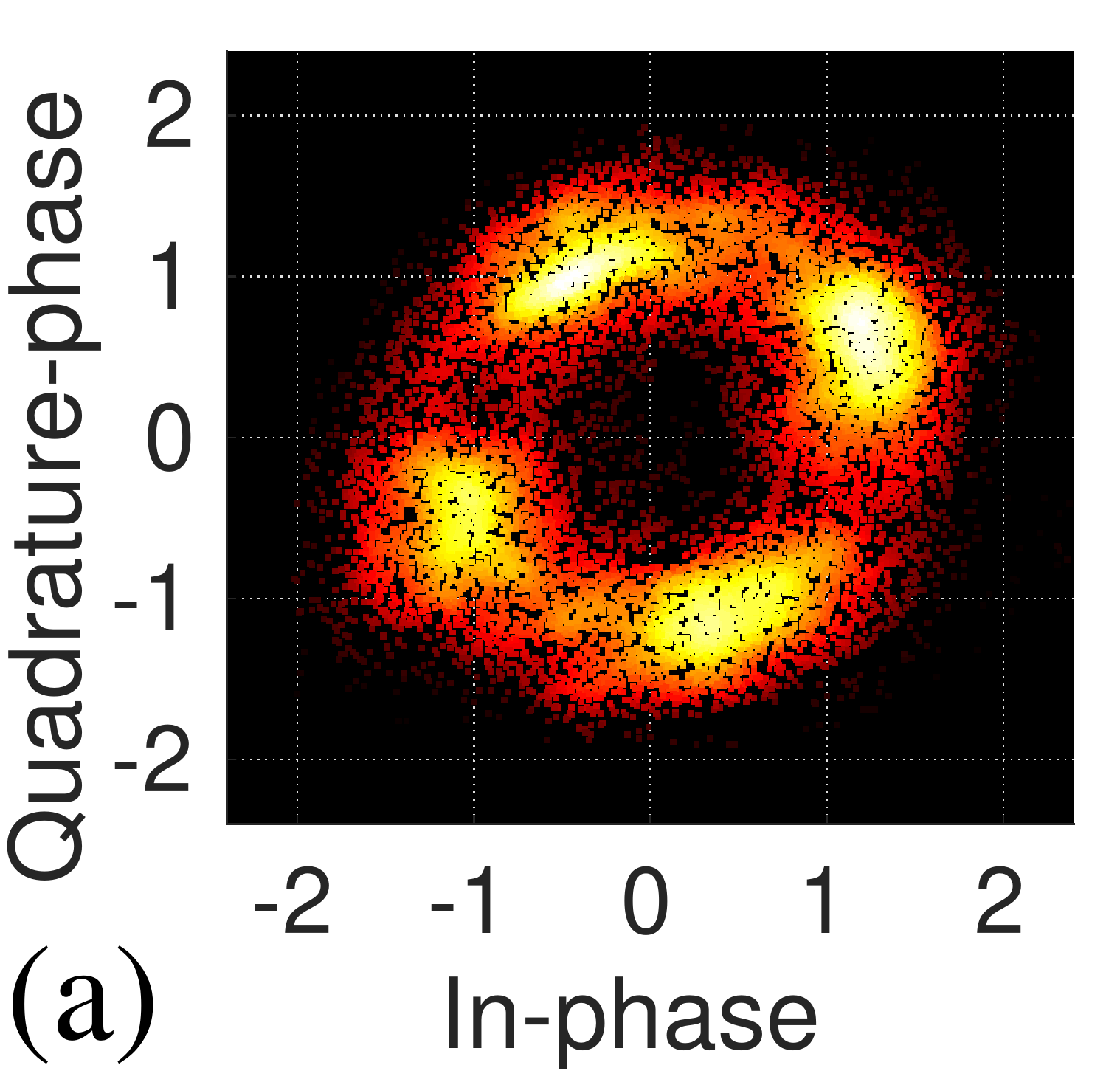}& 
		\hspace{-.05cm}\includegraphics[width=.245\textwidth]{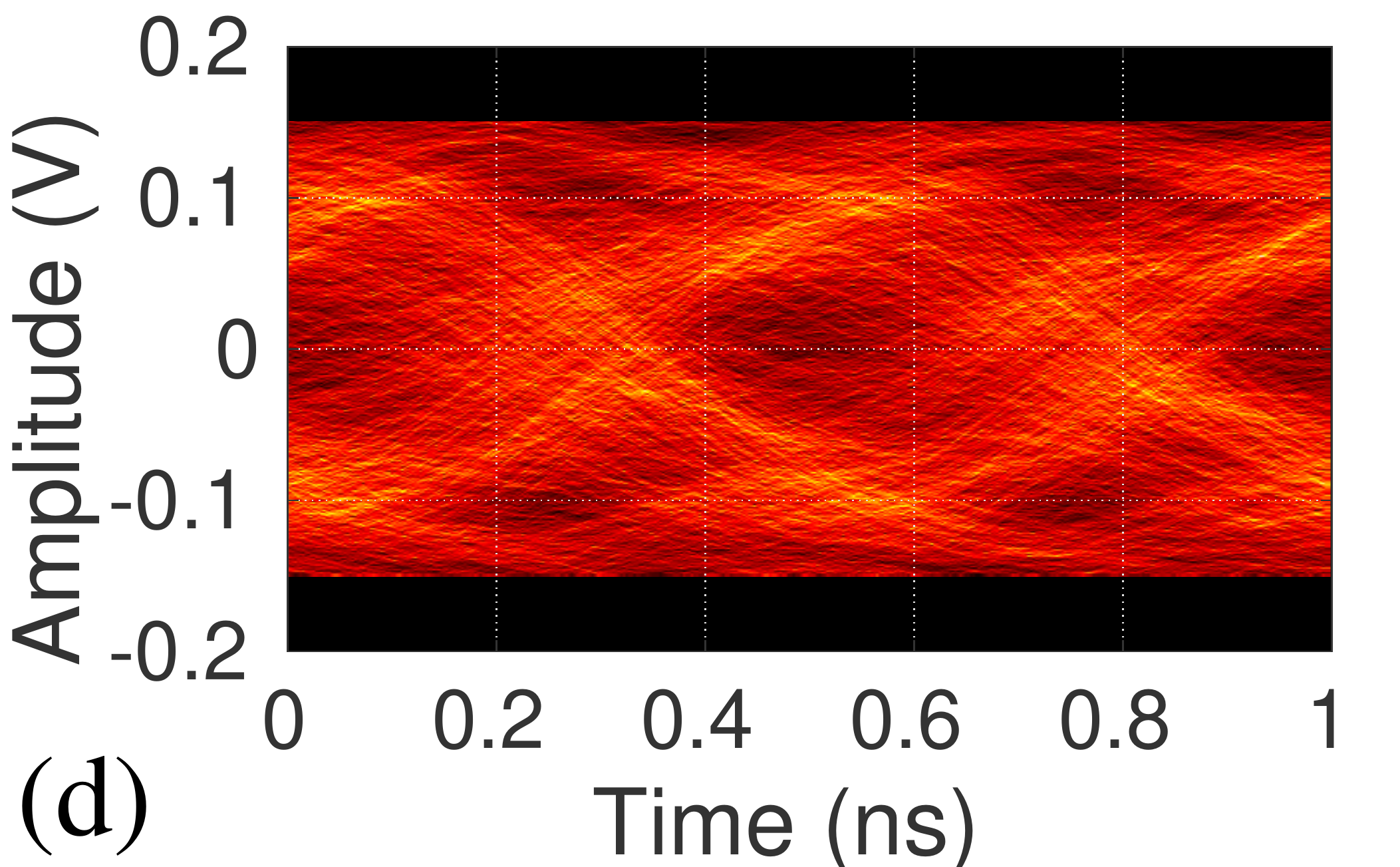}\\
		\hspace{-.05cm}\includegraphics[width=.16\textwidth]{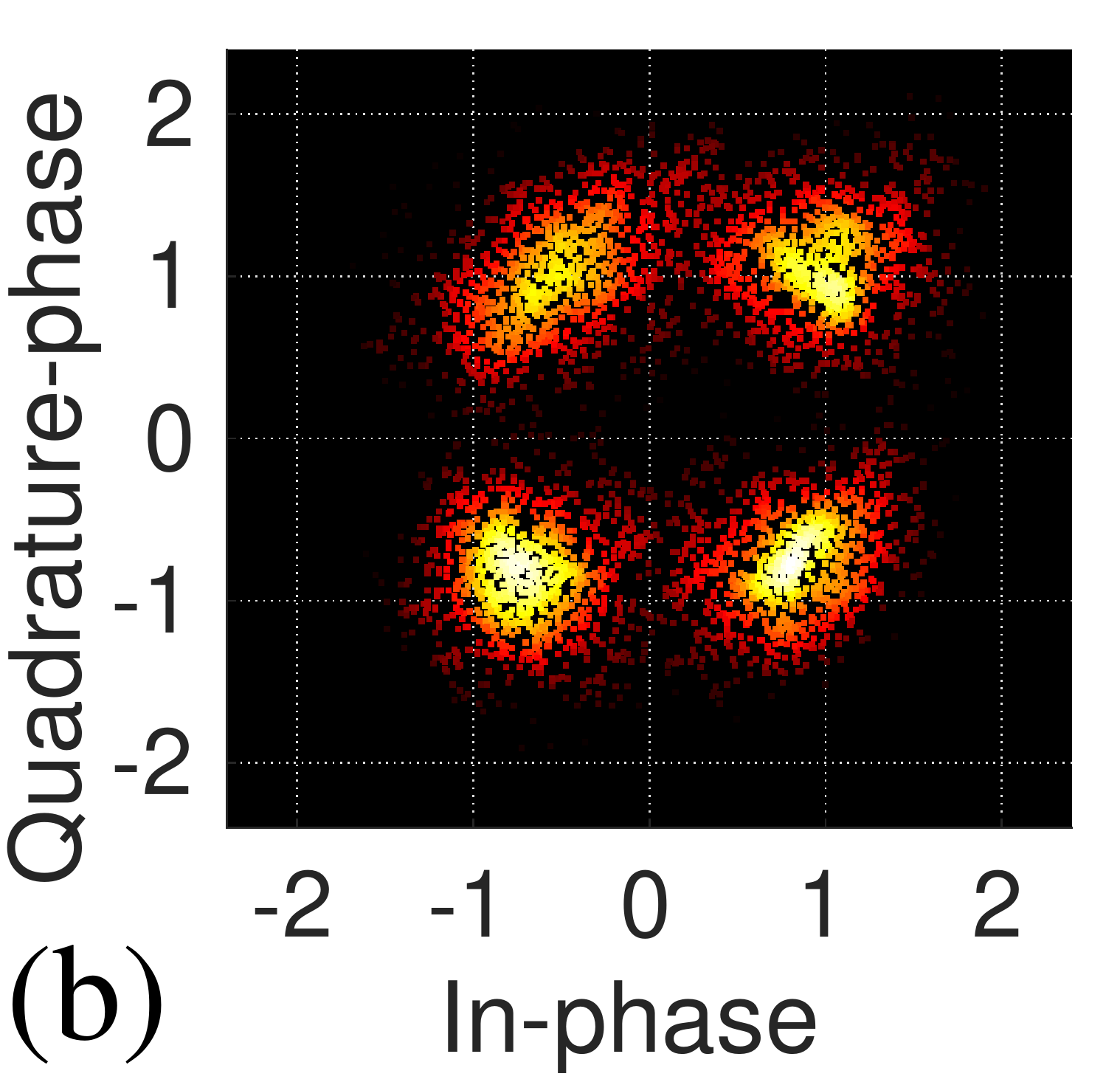}& 
		\hspace{-.05cm}\includegraphics[width=.245\textwidth]{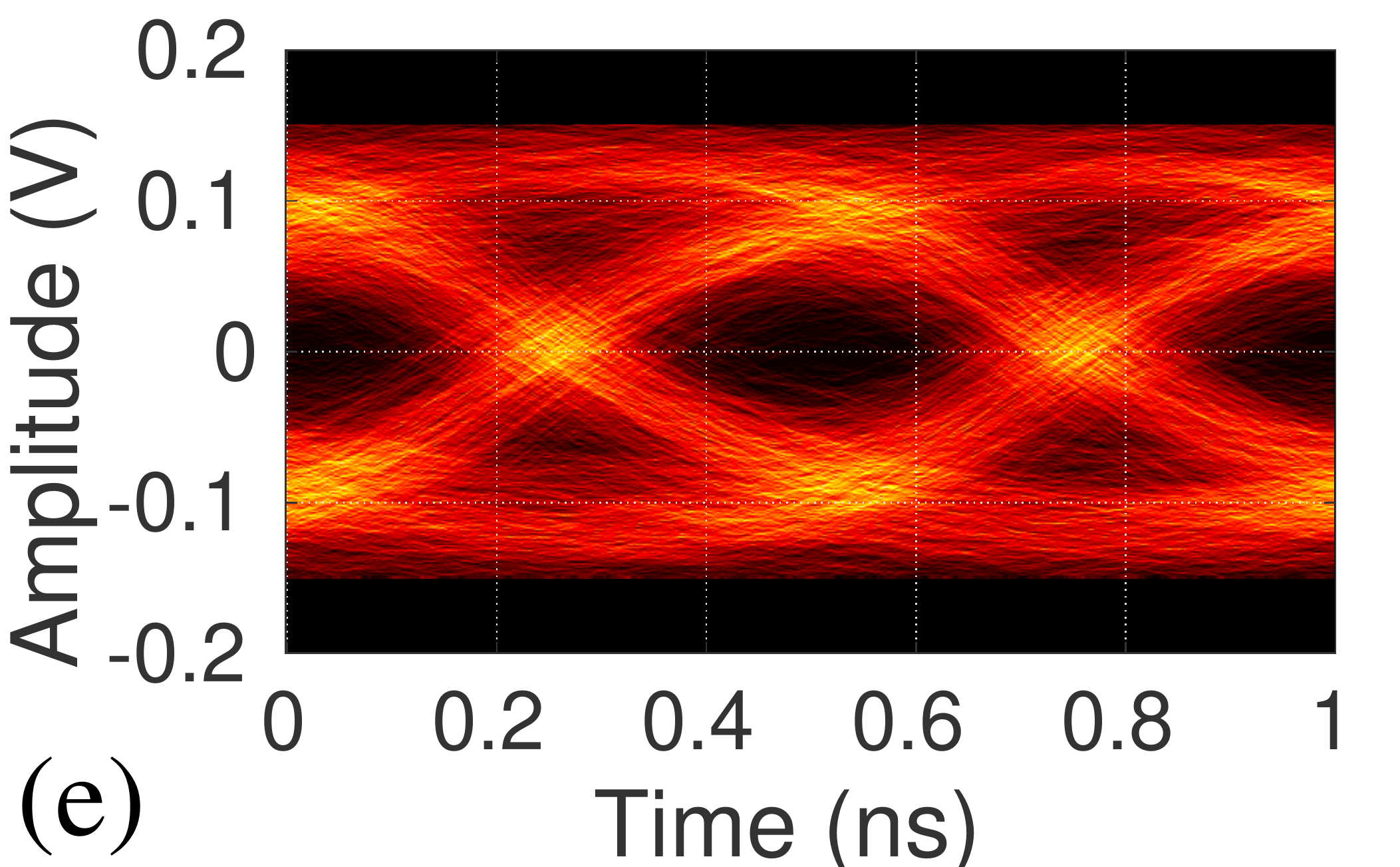}\\
		\hspace{-.05cm}\includegraphics[width=.16\textwidth]{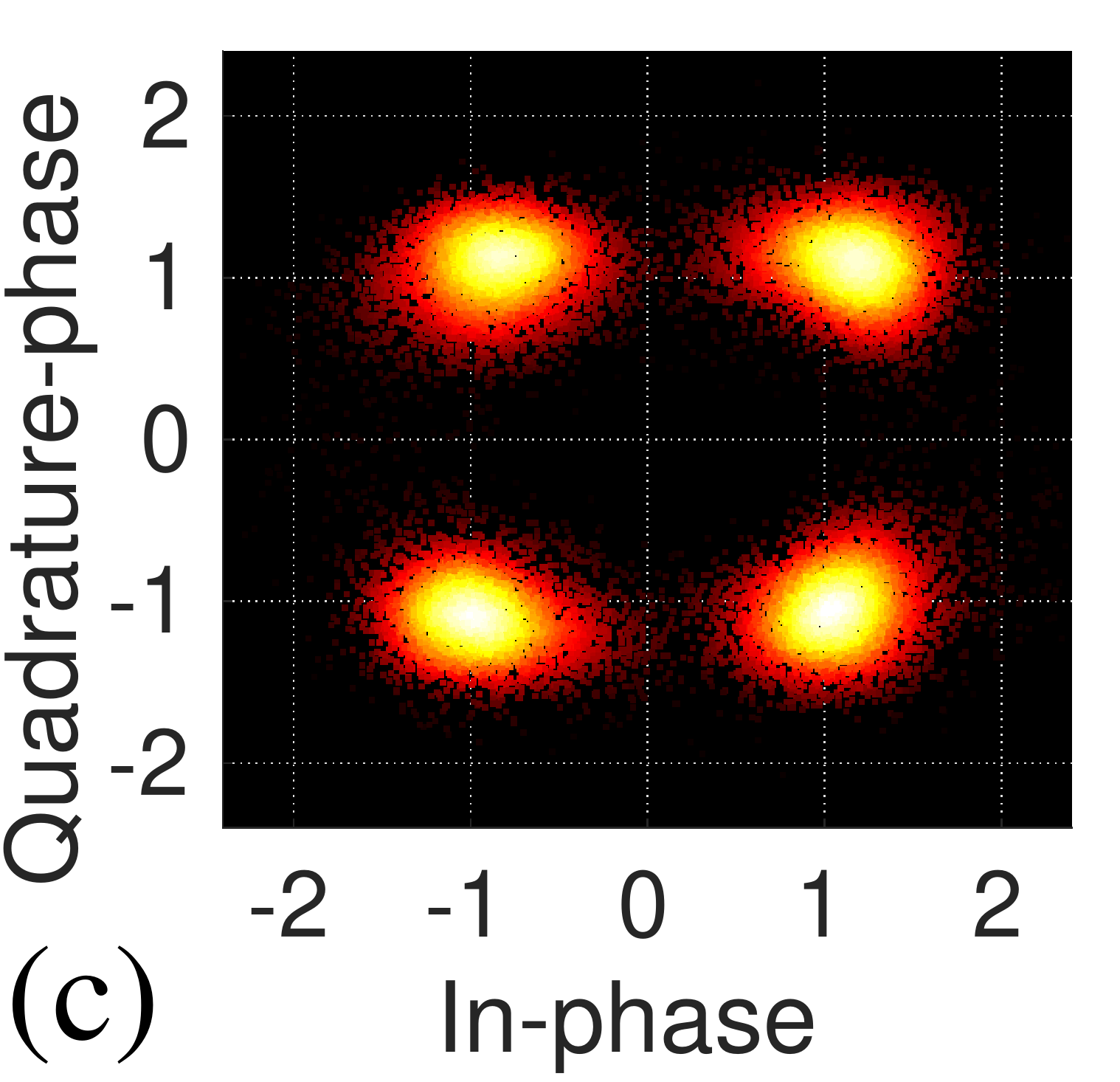}& 
		\hspace{-.05cm}\includegraphics[width=.245\textwidth]{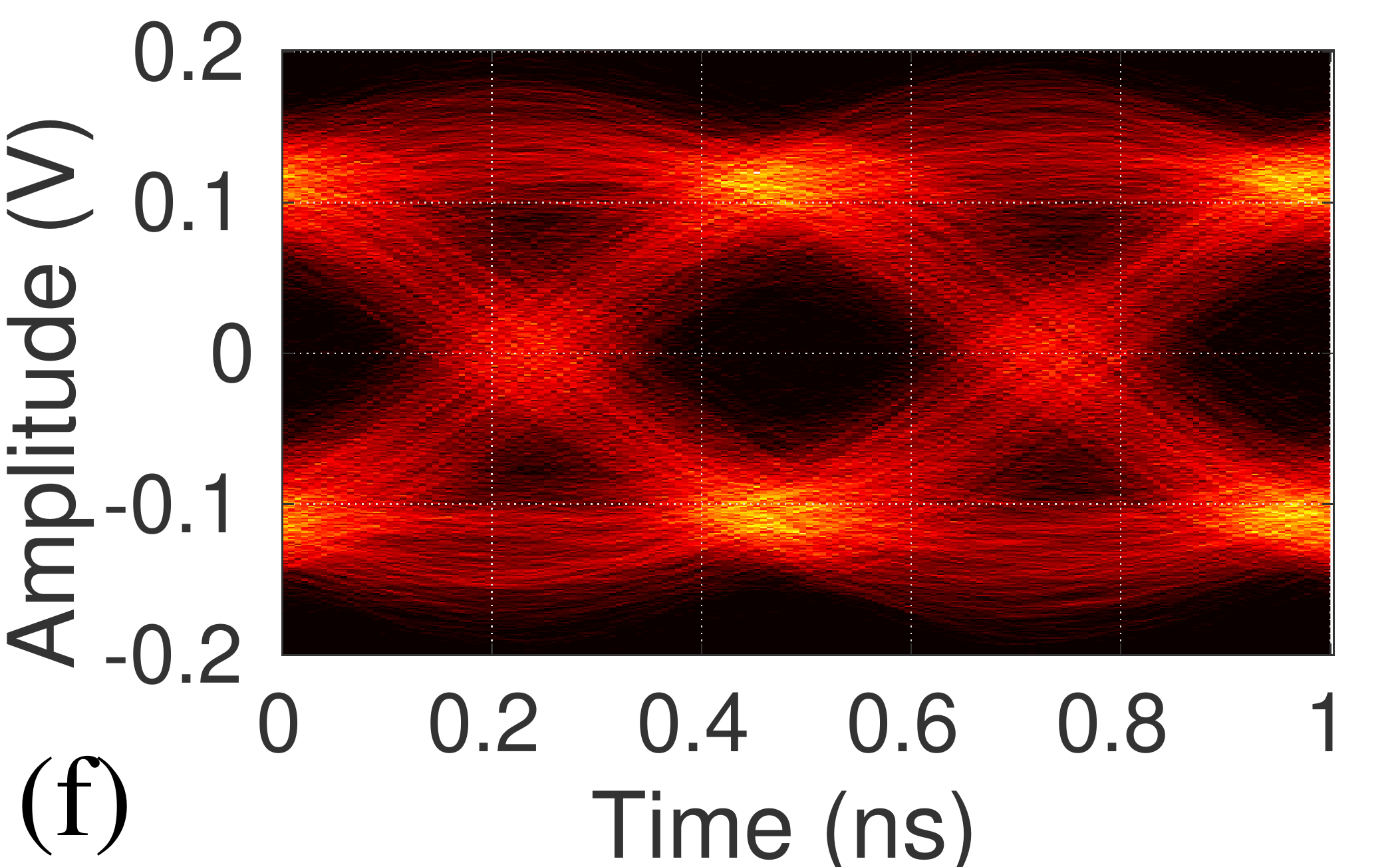}\\
\end{tabular}}
		\caption{ Experimental results obtained with  QPSK link. 
		Constellation diagrams for: (a) Open loop CPR chip output; (b) Closed loop CPR chip output; and (c) 
		Equalizer (post-processed) output. Eye-diagram of quadrature-phase component of: 
		(d) Open loop CPR chip output; (e) Closed loop CPR chip output; and (f) 
		Equalizer (post-processed) output.}		
		\label{fig:SiPhCPRC_results}
	\end{figure}
	
	
Figure \ref{fig:SiPhCPRC_results} presents the constellation and eye-diagrams of experimental results obtained with the proposed architecture. 
The open loop constellation points shown in Fig. \ref{fig:SiPhCPRC_results}(a) rotate with time due to continuously varying phase offset between $S$ and 
$LO$. The constellation settles when the receiver loop is closed as the phase offset between $S$ and $LO$ is corrected, as shown in Fig. \ref{fig:SiPhCPRC_results}(b).
 The closed loop signals are equalized to compensate for the frequency dependent losses and hence reduce the error vector magnitude through 
 post-processing, as shown in Fig. \ref{fig:SiPhCPRC_results}(c), which are further used for data recovery. 
 Eye diagrams of the measured quadrature-phase signals are presented in Figs. \ref{fig:SiPhCPRC_results}(d)--(f) and similar 
 plots can be obtained for in-phase signals too. The eye is closed in open loop condition (Fig. \ref{fig:SiPhCPRC_results}(d)) due 
 to time varying phase offsets, 
 while it gets opened in the closed loop condition (Fig. \ref{fig:SiPhCPRC_results}(e)) after the phase offset correction. 
 Horizontal and vertical eye openings are 
 increased post-equalization, as shown in Fig. \ref{fig:SiPhCPRC_results}(f).
 Single stage analog domain equalizers \cite{Nambath_JLT2020,Verplaetse_JSSC2020} or their cascade combination 
 \cite{Kamran_OFT2019} can be used for the equalization.

The data rate of measurement was constrained by the limited constant gain frequency range of the amplifiers connecting the ICR and baluns. 
The baluns made up of transformers are bandwidth limited. In addition, 
the chips are wire bonded to the PCBs, which have inherent parasitic inductances, limiting the frequency of operation.
These factors restrain the data rate of the demonstration even though the SiPIC ICR and SiGe CPR are designed to operate for 50\,Gbaud and 25\,Gbaud, 
respectively. Although both the chips are compatible with 16-QAM modulation scheme, signal to noise ratio (SNR) of the PIC restricts the operation of the chip to QPSK modulation format. 
With the aid of linear TIAs (currently not shown in Fig. \ref{fig:Exp_EPIC}(a)) at the output of ICR, 
the cumulative SNR of the system can be improved and hence higher order modulation formats and data rate can be 
attained.
The data rate of the assembly can be enhanced by using high gain wide frequency range amplifiers, broadband baluns, efficient optical coupling, and flip-chip bonded ICs. 
In addition, the overall baud rate of the EIC can be enhanced by designing the circuits in advanced CMOS/FinFET technology nodes.


\section{Conclusion}
We have presented an EIC-PIC codesigned analog coherent receiver solution for near-future DCIs. 
The proof-of-concept demonstration results of SiPIC ICR integrated with CPR chip at system level for QPSK modulation scheme are promising.
Since both the ICs are designed to function with 
16-QAM for higher data rates, the throughput per wavelength per polarization per channel can be increased by system level optimization. 
Electronic and photonic integration on a single chip or in a single package can result in compact and energy efficient solutions for high-speed data center interconnects.

\section{Acknowledgment}
The authors would like to thank Rashmi Kamran for the technical discussions. The authors would also like to thank the Ministry
of Electronics and Information Technology, Government of
India, for funding.

\bibliographystyle{IEEEtran}
\bibliography{references_v1}

\begin{thebibliography}{10}
\providecommand{\url}[1]{#1}
\csname url@samestyle\endcsname
\providecommand{\newblock}{\relax}
\providecommand{\bibinfo}[2]{#2}
\providecommand{\BIBentrySTDinterwordspacing}{\spaceskip=0pt\relax}
\providecommand{\BIBentryALTinterwordstretchfactor}{4}
\providecommand{\BIBentryALTinterwordspacing}{\spaceskip=\fontdimen2\font plus
\BIBentryALTinterwordstretchfactor\fontdimen3\font minus
  \fontdimen4\font\relax}
\providecommand{\BIBforeignlanguage}[2]{{%
\expandafter\ifx\csname l@#1\endcsname\relax
\typeout{** WARNING: IEEEtran.bst: No hyphenation pattern has been}%
\typeout{** loaded for the language `#1'. Using the pattern for}%
\typeout{** the default language instead.}%
\else
\language=\csname l@#1\endcsname
\fi
#2}}
\providecommand{\BIBdecl}{\relax}
\BIBdecl

\bibitem{Nambath_JLT2020}
N.~{Nambath} \emph{et~al.}, ``{All-Analog Adaptive Equalizer for Coherent Data
  Center Interconnects},'' \emph{J. Lightw. Technol.}, vol.~38, no.~21, pp.
  5867--5874, 2020.

\bibitem{Perin_JLT2021}
J.~K. {Perin}, A.~{Shastri}, and J.~M. {Kahn}, ``{Coherent Data Center
  Links},'' \emph{J. Lightw. Technol.}, vol.~39, no.~3, pp. 730--741, 2021.

\bibitem{Hirokawa_JLT2021}
T.~Hirokawa \emph{et~al.}, ``{Analog Coherent Detection for Energy Efficient
  Intra-Data Center Links at 200 Gbps Per Wavelength},'' \emph{J. Lightw.
  Technol.}, vol.~39, no.~2, pp. 520--531, 2021.

\bibitem{Kikuchi_JLT2016}
K.~Kikuchi, ``{Fundamentals of Coherent Optical Fiber Communications},''
  \emph{J. Lightw. Technol.}, vol.~34, no.~1, pp. 157--179, Jan 2016.

\bibitem{Ashok_JLT2021}
R.~Ashok, S.~Naaz, R.~Kamran, and S.~Gupta, ``Analog domain carrier phase
  synchronization in coherent homodyne data center interconnects,'' \emph{J.
  Lightw. Technol.}, vol.~39, no.~19, pp. 6204--6214, 2021.

\bibitem{Verplaetse_JSSC2020}
M.~Verplaetse \emph{et~al.}, ``{Analog I/Q FIR Filter in 55-nm SiGe BiCMOS for
  16-QAM Optical Communications at 112 Gb/s},'' \emph{IEEE J. Solid-State
  Circuits}, vol.~55, no.~7, pp. 1935--1945, 2020.

\bibitem{Ashok_JLT2021_1}
R.~Ashok, N.~Nambath, and S.~Gupta, ``Carrier phase recovery and compensation
  in analog signal processing based coherent receivers,'' \emph{Journal of
  Lightwave Technology}, pp. 1--1, 2021.

\bibitem{Jalali_JLT2006}
B.~{Jalali} and S.~{Fathpour}, ``Silicon photonics,'' \emph{J. Lightw.
  Technol.}, vol.~24, no.~12, pp. 4600--4615, 2006.

\bibitem{Ahmed_JSSC2020}
A.~H. Ahmed, A.~E. Moznine, D.~Lim, Y.~Ma, A.~Rylyakov, and S.~Shekhar, ``{A
  Dual-Polarization Silicon-Photonic Coherent Transmitter Supporting 552
  Gb/s/wavelength},'' \emph{IEEE J. Solid-State Circuits}, vol.~55, no.~9, pp.
  2597--2608, 2020.

\bibitem{Temporiti_ISSC2016}
E.~Temporiti, G.~Minoia, M.~Repossi, D.~Baldi, A.~Ghilioni, and F.~Svelto, ``{A
  56Gb/s 300mW silicon-photonics transmitter in 3D-integrated PIC25G and 55nm
  BiCMOS technologies},'' in \emph{Proc. ISSC}, 2016, pp. 404--405.

\bibitem{Doerr_JLT2010}
C.~R. Doerr \emph{et~al.}, ``{Monolithic Polarization and Phase Diversity
  Coherent Receiver in Silicon},'' \emph{J. Lightw. Technol.}, vol.~28, no.~4,
  pp. 520--525, Feb 2010.

\bibitem{Tsunashima_OE2012}
S.~Tsunashima \emph{et~al.}, ``{Silica-based, compact and
  variable-optical-attenuator integrated coherent receiver with stable
  optoelectronic coupling system},'' \emph{Opt. Express}, vol.~20, no.~24, pp.
  27\,174--27\,179, Nov 2012.

\bibitem{Valdecasa_TCAS22021}
G.~S. Valdecasa, O.~G. Puertas, J.~A. Altabas, M.~Squartecchia, J.~B. Jensen,
  and T.~K. Johansen, ``{High-Speed SiGe BiCMOS Detector Enabling a 28 Gbps
  Quasi-Coherent Optical Receiver},'' \emph{IEEE Transactions on Circuits and
  Systems II: Express Briefs}, pp. 1--1, 2021.

\bibitem{Kamran_JLT2020}
R.~{Kamran}, S.~{Naaz}, S.~{Goyal}, and S.~{Gupta}, ``{High-Capacity Coherent
  DCIs Using Pol-Muxed Carrier and LO-Less Receiver},'' \emph{J. Lightw.
  Technol.}, vol.~38, no.~13, pp. 3461--3468, 2020.

\bibitem{lockwood2010silicon}
D.~J. Lockwood and L.~Pavesi, \emph{{Silicon photonics II: Components and
  integration}}.\hskip 1em plus 0.5em minus 0.4em\relax Springer Science \&
  Business Media, 2010, vol. 119.

\bibitem{Ashok_OFC2019}
R.~{Ashok}, S.~{Manikandan}, S.~{Chugh}, S.~{Goyal}, R.~{Kamran}, and
  S.~{Gupta}, ``{Demonstration of an Analogue Domain Processing IC for Carrier
  Phase Recovery and Compensation in Coherent Links},'' in \emph{Proc. OFC.},
  2019, pp. 1--3.

\bibitem{Ashok_CLEO2020}
R.~Ashok, R.~Kamran, S.~Naaz, and S.~Gupta, ``{Demonstration of a PMC-SH link
  using a phase recovery IC for low-power high-capacity DCIs},'' in
  \emph{Conference on Lasers and Electro-Optics}, 2020, p. SF3L.3.

\bibitem{Kamran_OFT2019}
R.~Kamran, S.~Manikandan, and S.~Gupta, ``{Cascaded equalizer for polarization
  multiplexed carrier based self-homodyne QPSK links},'' \emph{OFT}, vol.~50,
  pp. 233--241, 2019.

\end{thebibliography}

\end{document}